# The Hot Mitochondrion Paradox: Reconciling Theory and Experiment

Peyman Fahimi,[a,b] Chérif F. Matta,*[a-d]

[a] *Department of Chemistry and Physics, Mount Saint Vincent University, Halifax, Nova Scotia, Canada B3M2J6*

[b] *Département de chimie, Université Laval, Québec, Québec, Canada G1V 0A6*

[c] *Department of Chemistry, Saint Mary's University, Halifax, Nova Scotia, Canada B3H3C3*

[d] *Department of Chemistry, Dalhousie University, Halifax, Nova Scotia, Canada B3H4J3*

* Correspondence: cherif.matta@msvu.ca (C. Matta)

**Abstract**

Experiments by Chrétien and co-workers suggest that mitochondria are 10$^{o}$C hotter than their surroundings. Steady-state theoretical estimates place this difference at a maximum of $10^{-5}$ $^{o}$C. This million-fold disagreement may be called the "*hot mitochondrion paradox*". It is suggested that *every* proton translocated via ATP synthase sparks a picosecond temperature-difference spike of the order of magnitude measured by Chrétien *et al.* Time-averaging of these spikes recovers the theoretical value. Further, a temporal and spatial superposition of the fluorescence intensity of a very large number of molecular thermometer molecules in the sample can give the appearance of a steady signal. The inner mitochondrial membrane appears to be flanked by temperature differences fluctuating in time and along the membrane's surface, with "hot" and "cold" spots as ultrashort temperature spikes.

**An apparent millionfold disagreement of theory and experiment**

The mitochondrion as the "powerhouse of the cell" is a phrase of universal use that was originally coined by Philip Siekevitz in 1957 [1]. Mitochondria generate the majority of the energy supply of most eukaryotic cells whereby the energy in foodstuff is transduced in this fascinating organelle into the universal energy currency of life; namely, ATP. The enzyme responsible for the coupling of the mitochondrial proton gradient created by the continual release of energy from the electron transport chain (ETC) is **ATP synthase** (see Glossary) (Box 1). This enzyme, a molecular rotary machine, is the focus of this opinion article in which we show how it is responsible not only for the production of ATP but also for the maintenance an oscillatory temperature gradient within the mitochondrion.

Fluorescent thermometry is an important tool used to elucidate temperature variations on the nano-scale in living cells (Box 2). In 2018, Chrétien and colleagues [2], using a fluorescent probe that accumulates into mitochondria [*mito thermo yellow* (MTY)] [3], proposed that mitochondria are considerably "hotter" than their immediate surrounding by up to 10$^{o}$C. This extraordinary claim has been received with skepticism [4] and considerable attention and has, so far, survived rigorous and extensive vetting [5–7]. The claim is puzzling since a temperature difference of a few degrees over just a few nanometers would entail an unprecedented temperature gradient across the mitochondrial membrane structure. Meanwhile these claims are corroborated by the unusually high expression of heat shock proteins in mitochondria, possibly to protect its critical macromolecules from melting and from of reactive oxygen species damage [8].

The enormous temperature gradient implies an *extremely* low thermal conductivity that can hardly be consistent with a double phospholipid bilayer membrane. Steady-state modeling of heat transfer in mitochondria predicts temperature differences that are several orders of magnitudes lower. A **paradox** then arises, since Chrétien and colleagues' experimental results themselves (regardless of their interpretation) were carefully conducted and never refuted. What could be questioned, however, is the *interpretation* of the experimental results on the one hand and the steady-state assumption on the other.

The *Hot Mitochondrion Paradox*, as we would call it, can be restated in two questions: (*i*) How can the observed fluorescence intensities of MTY be converted into local mitochondrial temperatures? and (*ii*) Could it be hypothesized that there exist *localized regions* with temperatures that are *transiently* higher than the surrounding in mitochondria ("hot spots"/"hot flashes", respectively)?

Using the Fourier Law of heat diffusion (Box 3), Baffou and colleagues [9] showed that a steady-state temperature difference in a single cell cannot exceed from $10^{-5}$ K,

$$\Delta T = \frac{P}{\kappa L}, \quad (1)$$

where $P, \kappa,$ and $L$ are, respectively, the cellular power (total flow of energy per unit time required to sustain the life of the cell), the thermal conductivity of the watery environment, and the cell size. This $\Delta T \approx 10^{-5}$ K is obtained using the consensus values $P$ = 100 pW, $L$ = 10 µm, and $\kappa$ = 1 W $m^{-1}$ $K^{-1}$ (for comparison, for pure water $\kappa \approx 0.66$ W $m^{-1}$ $K^{-1}$ [10]). This is came to be known as "***the*** $\boldsymbol{10^5}$ ***gap***", which is supported by other theoretical analyses [11,12].

Meanwhile, using a nanothermometer hybrid, Sotoma and colleagues [13] found that both MCF-7 and HeLa cells have a thermal conductivity of 0.11 ± 0.04 W $m^{-1}$ $K^{-1}$. This thermal conductivity is an order of magnitude lower than that used above, changing the gap to $10^{-4}$ K, which is still far from the 10°C reports.

Macherel and colleagues [12] explore different possibilities to justify the 10°C difference. They conclude that the main mechanism of heat transfer in the cell is *diffusion* and that the characteristics of the mitochondrial membrane structure cannot explain the low thermal conductivity necessary to maintain such a gradient. These authors [12] argue that physical models appear incapable to account for the puzzle of the "hot mitochondrion" despite similar findings with respect to the temperature of the mitochondrion and of the cell nucleus [14–17].

**Temperature difference and proton translocation through ATP synthase**

At physiological pH, the MTY "thermometer" (Box 2) is believed to be mainly distributed in the mitochondrial matrix side of the inner membrane and possibly in the membrane bound to ALDH2 [5].

Basal proton leak/slippage, which is responsible for at least 20% of the basal metabolic rate [18–20], plays a dominant role in mitochondrial heat generation [21]. Basal leakage appears to be mediated by "water wires" in the lipid bilayers [19,20], adenine nucleotide translocase (ANT) [19,20], or ATP synthase [22–26].

A proton is, in essence, a "heat carrier" as it trickles down the electrochemical gradient releasing heat as it emerges from the matrix side irrespective of its coupling to ATP production (*vide infra*). The power per ATP synthase – that is, the rate of Gibbs energy released during proton translocation – is:

$$\Delta\dot{G} = \dot{n}F(\mathrm{PMF}), \tag{2}$$

where $\dot{n}, F,$ and (PMF) are the number of protons per second passing through ATP synthase, the Faraday constant, and the proton-motive force, respectively. Given that the dominant contribution to $\Delta G$ in Eq. (2) is enthalpic (≈ 90%), the entropic contribution ($T\Delta S$) can hence be ignored in a first analysis [27,28]; that is, $\Delta H \approx \Delta G$ is assumed.

*Every proton translocation* through ATP synthase, irrespective of its coupling with ATP synthesis, contributes to the temperature difference. The rate of heat generation that accompanies proton transfers (leak and ATP-coupled transfers), the protonation of water by the proton emerging into the matrix, and the subsequent hydration of the ensuing hydronium ion, will be shown to account for the claimed 10$^{o}$C *only* when the mechanism of heat transfer is considered. The reported high temperature difference is strongly dependent on the *mechanism* of proton translocation. Among the different ways that protons can translocate, emphasis is placed here on proton translocation through ATP synthase since this enzyme's rotary mechanism will be shown to be crucial in understanding its thermal conduction.

**Thermal conductance of ATP synthase as a ratchet engine**

Resolving an apparent inconsistency regarding **Feynman's ratchet engine** [29], Parrondo and Español [30] emphasize that *the thermal conductance of a ratchet engine cannot be explained by considering the thermal conductivity of its constitutive material*. In doing so, they obtained the thermal conductance of an **axle-vane engine** through the physical parameters that permit the axle-vane to work; note that this is above and beyond the thermal conduction of the axle itself [30].

Parrondo and Español modify Feynman's original irreversible engine by considering a non-rigid axle (that can sustain torsion) with two identical vanes at its two ends [30]. The vanes are kept into two isolated heat baths with a temperature differential (Figure 1). The system, depicted in Figure 1, has two degrees of freedom represented by the vanes angles, which are partially coupled. These angles fluctuate under the random bombardment due to the thermal motion of the molecules in each of the two baths. When the combined system reaches a stationary state, these authors show that the thermal conductance of the axle has the following form [30]:

$$\frac{\dot{Q}}{\Delta T} = \kappa' = \frac{\lambda k_{\mathrm{B}} \alpha}{2m(1+\alpha)}, \qquad (3)$$

in which $\alpha = \frac{\tau m}{\lambda^2}$ is a dimensionless parameter that determines the stiffness of the axle, $\lambda$ $\left(\frac{\mathrm{N.s}}{\mathrm{m}}\right)$ is the friction coefficient of vanes with the environment, $\tau$ $\left(\frac{\mathrm{N}}{\mathrm{m}}\right)$ is the torsion coefficient, and $m$ is the mass of each one of the two vanes. Eq. (3) has been also reached by other authors [31,32].

When thermal fluctuations lead to the motion of a motor which works through rotational degrees of freedom, the motor can be considered as a ratchet [33]. That is why the axle-vane system of Parrondo and Español can be considered as a ratchet engine.

The well-known ATP synthase rotary motor [34–37] has been investigated as a Feynman's ratchet-like engine in the literature [33,38–40]. For example, Oster and Wang [33,41] showed that the $F_O$ motor can play the role of a ratchet. These authors argue that, as the proton hops from one acidic residue to the next inside the $F_O$ channel, the continual on/off switching of the ensuing electrostatic forces favors the drift in one rotatory direction overwhelming the random Brownian rotations that tend to average to zero over time [42]. In this sense, Oster and Wang describe the $F_O$ unit as taking "*a hybrid*" quality "*with characteristics of both a Brownian ratchet and a power stroke*" [41].

Wang and Oster found an equation similar to Eq. (3) for the hydrolysis mode of ATPase in the absence of external torque [33].

By accounting for thermal angle fluctuations of the $\gamma$-subunit, the friction and torsion are identified as the two essential parameters governing this rotation [43–45]. The friction coefficient becomes important at rotational speeds exceeding $2^{o}$/ns [45]. Okazaki and Hummer [45] modeled the $\gamma$–rotation in synthesis mode of $F_1$-ATP synthase in such a manner that the $\gamma$–subunit is coupled to the $\alpha_3\beta_3$ subunits by a spring with two torsional constants, one for the twisted part and one for the connecting regions (Figure 2). In Okazaki and Hummer's model, reproduced with modification in Figure 2 (*right*), the $\gamma$-subunit is also connected to the *c*-rings in $F_O$ region but without a spring between them [45]. One can thus assume that the $\gamma$-subunit and the *c*-rings rotate with the same angular velocity.

As shown in Figure 2, domains 1 and 2 have different diffusion constants and, through the relation $\lambda = \frac{k_B T}{D}$, different friction coefficients with their immediate environment. The rotational diffusion and friction coefficients characterize the rotation of the *c*-rings-γ-subunit taken together in a primarily hydrophobic membrane environment and the $\alpha_3\beta_3$-subunit in the aqueous environment of the matrix.

Molecular dynamics (MD) simulation of an idealized ATP synthase by Okazaki and Hummer [45] yields an effective spring constant (torsion coefficient) $\tau_{\text{eff}} = \left(\frac{1}{\tau_1} + \frac{1}{\tau_{12}}\right)^{-1} \approx 460\ \text{pN.nm/rad}^2$ with a standard error of $\approx 280\ \text{pN.nm/rad}^2$ [45]. Okuno and colleagues [44] measured the torsional constant of the embedded part of the $\gamma$-subunit within the $\alpha_3\beta_3$-subunit with a magnetic bead using the thermophilic *Bacillus* PS3. These workers report a torsional stiffness of 223±141 pN.nm/rad$^2$ [44], which is within the error bars of the Okazaki and Hummer. This agreement of modeling and experiment mutually reinforces confidence in both.

The diffusion coefficient of the core domain in $F_1$ ($D_1$), taken from the trajectories of angle – that is, the temporal evolution of a given angle, is calculated by Okazaki and Hummer [45] to be equal to $\approx\ 0.001\ \text{rad}^2/\text{ns}$. This latter value is about 70 times less than the hydrodynamical diffusion of the free $\gamma$-subunit in aqueous solution and thus the $\gamma$-subunit in ATP synthase experiences considerably high friction [45].

The spherical and the cylindrical parts of Figure 2 can be identified with the two vanes of Parrondo and Español displayed in Figure 1. The torsional spring, in this analogy, plays the role of the axle in Parrondo and Español's model [30]. Notice that the "axle" subunit of the ATP synthase enzyme (Figure 2, left), as is conventionally known, is *not* the same as the axle in Figure 1.

A Parrondo and Español-type double vane engine model that captures the essential physics of thermal conductance of ATP synthase is encasulated in Eq. (3). The rotary degree of freedom of ATP synthase is governed by friction and torsional forces, which can hamper the rotary motion. If torsion and friction could stop this motion, ATP synthase becomes not only incapable of catalyzing ATP synthesis (or splitting) but not even allowing its possible contribution to proton slippage. That would necessarily imply the stoppage of ATP production and ATP synthase-mediated heat production.

Parrondo and Español derive the thermal conductance of the axle by considering the *linear motion* of the blades of the vanes [30]. Meanwhile, the results by Okazaki and Hummer [45] provide estimates of parameters in Eq. (3) for ATP synthase (*vide infra*) but expressed in terms of a *rotary motion*. Given that we have only the rotational expression of these parameters [46], one needs to convert Parrondo and Español's equation into its rotational form. Hence, we use the moment of inertia (instead of mass), the rotational diffusion (instead of the translational one), and the angular form of Hooke's law for the torsional spring. Box 4 reviews some key concepts related to the units and dimensions involving rotational motion.

Inserting Eq. (2) in the rotational form of Eq. (3) and considering the rate of heat transfer from the cylindrical domain toward the spherical domain (Figure 2) gives

$$\Delta T = \frac{2\,\dot{n}\,F\,(\mathrm{PMF})\,I_{\gamma+\mathrm{c}}\,(1+\alpha)}{\lambda\,k_{\mathrm{B}}\,\alpha}, \tag{4}$$

where $\lambda = \frac{k_B T}{D_1}\left[\frac{\text{J.s}}{\text{rad}^2}\right]$, $\alpha = \frac{\tau_{\text{eff}} I_{\gamma+c}}{\lambda^2}$, and $I_{\gamma+c} = \frac{1}{2}\left(M_\gamma R_\gamma^2 + M_c R_c^2\right)\left[\frac{\text{kg.m}^2}{\text{rad}^2}\right]$ is the moment of inertia of the $\gamma$- and $c$-subunit as two attached solid cylinders with masses $M_\gamma, M_c$ and radii $R_\gamma, R_c$ per radian (rad) (Figure 2 and Box 4) around the $z$-axis.

Underlying Eq. (4) is the approximation $\Delta\dot{G} \approx \Delta\dot{Q}$, since the enthalpic term represents 90% of the $\Delta G$ of ATP synthesis [27,28]. The mass of the gamma subunit ($M_\gamma$) is ≈ 30,337 Da[i] and that of the $c$-rings ($M_c$) of, say, eight $c$-subunits, is ≈ 8×14,200 Da[ii], and the rate of proton translocation through ATP synthase ($\dot{n}$) is roughly 1200 protons/s (or ≈ 1 proton/0.00083 s) [47]. With these numerical values, we obtain (in SI units) $\lambda \approx 4 \times 10^{-27}\left[\frac{\text{J.s}}{\text{rad}^2}\right]$ and $\alpha \approx 6.70 \times 10^{-6}$; hence, $\alpha \ll 1$ but *not* zero. The smallness of $\alpha$ implies that the axle (spring) is torsionally soft.

Inserting the explicit expression of $\alpha$ in the denominator of Eq. (4) and ignoring it in the numerator, the temperature difference expression between the entry and exit point of protons is

$$\Delta T \cong \frac{2\,\dot{n}\,F\,(\text{PMF})\,\lambda}{k_B\,\tau_{\text{eff}}}, \qquad (5)$$

which on substitution of typical values leads to ΔT ≈ 0.1 K. This estimate, while four orders of magnitude greater than the $10^5$ gap, is still far from the 10°C difference. We now propose a new idea to resolve this apparent conflict. A plausible way out of this conundrum is by accounting for the (de)protonation of water and the (de)hydration of the hydronium ions. These processes, since they occur in opposite orders on the two sides of the membrane, may be overlooked as they contribute little to the net $\Delta G$ averaged over timescales beyond a few dozens of nanoseconds. However, these mechanisms will be shown below to be crucial in generating the reported temperature difference despite of an almost zero contribution to the net power of the mitochondrion.

**Protonation, hydration, and thermalization**

Protons in both the intermembrane gap and in the mitochondrial matrix are present as solvated hydronium ions, exchanging on a sub-100-fs scale [48] between hydrogen-bonded water molecules. ATP synthase allows the passage of protons through its $F_O$ unit, but in which form: hydronium ion or free proton? This question has been worked out by the MD simulations of Leone,

Krah, and Faraldo-Gómez [49]. They find that, during transit within the $F_O$ unit, the proton is ping-ponged as such between acceptors in the channel and does not travel as hydronium ion [49]. When a $H^+$ emerges from the matrix side, it protonates a water molecule with a subsequent reorganization of the solvent shells around the newly formed hydronium ion.

There exist two reverse sequences of events, one on the intermembrane gap side and the other on the matrix side. At the intermembrane gap side, desolvation and deprotonation occur first prior to entry in ATP synthase as a "naked" proton. These steps are endergonic, *ca.* +418 kJ//mol (+100 kcal/mol) for desolvation and *ca.* +765 kJ/mol (+183 kcal/mol) for deprotonation, under standard conditions [50]. On exit of the proton on the matrix side, the two reverse processes are exergonic with similar magnitudes. There is thus an endergonic (two step) process on the gap side to be undone in reverse order by an exergonic one on the matrix side. Over time, energy balances-out on average save for the small variations due to local conditions on entry and exit.

Any proton undergoing translocation and emerging from the $F_O$ unit acquires the thermal bath energy, which is entirely kinetic, since a proton does not have any internal (electronic, vibrational, or rotational) degrees of freedom. This energy is then transmitted through collision to the first encountered water molecule and its hydrogen-bonded partners during the process of protonation. The protonated water, $H_3O^+$, is then solvated by the surrounding $H_2O$ molecules. This sequence of events is summarized in Figure 3.

Figure 3 provides estimates for the energies of protonation and solvation/hydration, -418 kJ/mol and -765 kJ/mol, respectively [50], while the timescale ($\tau_s$) of (de)hydration is in picoseconds [48,51] and that of (de)protonation is in femtoseconds [49]. Figure 3 also gives the temporal Heisenberg uncertainties for exchanging the corresponding energies ($\Delta G^0 \Delta\tau_s \gtrsim \frac{\hbar}{2}$), where $\tau_s >$ (or $\gg$) $\Delta\tau_s$.

The timescale for a translocated proton to collide with water is estimated as follows. The typical intermolecular distance in liquid water under normal conditions is *ca.* 0.27–0.30 nm [52]. The indeterminacy of the position of a newly released proton can thus be taken as ~ 0.3 nm. In the case of a proton leak, uncoupled to the synthesis of ATP, the chemiosmotic energy plus the thermal bath energy, $\Delta G^0_{\mathrm{CO+TB}} \approx 3.64 \times 10^{-20}$ J, is converted totally to kinetic energy resulting in an average proton speed of $\approx 6.6 \times 10^3$ m/s. Dividing 0.3 nm by $6.6 \times 10^3$ m/s gives the time scale $\tau_s \approx 5 \times 10^{-14}$ s for the period from proton exit until it collides with the first water molecule to transmit its $\Delta G^0_{\mathrm{CO+TB}}$ as kinetic energy to water. When chemiosmotic coupling uses the proton's $\Delta G^0_{\mathrm{CO}}$ in ATP synthesis, the proton has only the heat bath energy (at a temperature taken here as 25 °C) as the driver of its random motion. In such case, the characteristic time is longer by an order of magnitude; that is, $\approx 10^{-13}$ s (Figure 3).

Given the large difference in mass between a proton and ATP synthase ($\approx 1 : 6\times10^5$), we assume that all the kinetic energy is carried by the proton. This energy is the first to be released through collision with the water hydrogen-bonded network at a timescale $\tau_s \approx 5 \times 10^{-14}$ s (Figure 3), two to four orders of magnitude shorter than that of molecular vibrations ($10^{-12}$ - $10^{-10}$ s) [12,53] necessary to dissipate this energy.

The $\Delta G$ of protonation and hydration (on the matrix side), in addition to the much smaller proton translocation energy, are released on a timescale shorter than that necessary for the dissipation of energy through molecular translational, rotational, and vibration degrees of freedom. It appears reasonable, hence, to sum the $\Delta G$ (per proton) of these three steps (Figure 3) to re-estimate the $\Delta T$ from Eq. (5), now including all contributions, as follows:

$$\Delta T \cong \left\{ \frac{2\,\dot{n}\left[\Delta G^0_{\mathrm{Protonation}} + \Delta G^0_{\mathrm{Hydration}} + \Delta G^0_{\mathrm{Chemiosmotic\,+\,heat\,bath}}\right]\lambda}{k_B\,\tau_{\mathrm{eff}}} \right\}_{10^{-15}\mathrm{s} \le \tau_s \le 10^{-12}\mathrm{s}}, \quad (6)$$

where, on substitution of the numerical values as above, one obtains $\Delta T \approx 6.4$ K.

Once the total $\Delta G$ of three steps is released primarily as heat, heat diffusion **thermalizes** the microenvironment on a timescale of $10^{-12}$–$10^{-10}$ s. This range has been reached as follows. Thermalization in water occurs by a variety of mechanisms, including reorientation/rotation of water molecules, transfer of the vibrational excitation energy of a free (non-hydrogen bonded) O–H to a hydrogen-bonded arm of a water molecule (intermolecular energy transfer), and whole-molecule libration mediated via overtone of the bending mode of O–H [54]. Using a femtosecond two-color IR-pump/vibrational sum-frequency probe technique, Hsieh and colleagues report the timescale of thermalization for water molecules at the air/water interface by these mechanisms to be *ca.* 840±50 fs [54]. In another pump-probe terahertz spectroscopic study of water and deuterium oxide, Backus and colleagues reach a similar relaxation timescale for thermalization in aqueous media (*ca.* 0.6 ps) [55]. The hydrated proton dynamics is also in the hundreds of femtoseconds' timescale of vibrational relaxation [53]; in this case, 200-300 fs as revealed by an ultrafast IR spectroscopic study of acidic solutions by Carpenter and colleagues [56]. There can be an additional slower mechanism for the final thermalization including, for instance, the reorganization of the surrounding hydrogen-bonded networks, which occurs in a 2–3 ps timescale [56], and, finally, rotational relaxation, which happens at the 100-ps timescale [53].

In the thermalization step, there is no axle-vane mechanism and the heat diffuses by Fourier's law, which, through insertion in Eq. (1), $\Delta T \approx (2.0\times10^{-18}\text{J}/8.3\times10^{-4}\text{s}) / (0.1\ \text{Wm}^{-1}\text{K}^{-1}\times5\text{nm}) \approx 0.5\times10^{-5}$ K. Now time averaging over this temperature difference and that obtained from Eq. (6) yields $\Delta T_{\text{average}} \approx \{[(10^{12} - 1200)\times0.5\times10^{-5}\ \text{K}]+1200\times6.4\ \text{K}\}/10^{12} \approx 0.5\times10^{-5}$ K, where the average is identical to the instantaneous value within the given precision. This time averaging smears the temperature difference so that the picosecond spikes of $\Delta T \approx 6^{\circ}$C are flattened to recover the $10^5$ gap predicted by steady-state theoretical analysis [9,12].

An idealized representation of the temperature difference $\Delta T$ as a function of time on the basis of Figure 3 and Eq. (6) is depicted in Figure 4. Figure 4a illustrates the peaks resulting from the fusion of the three steps in Figure 4b into one large $\Delta T$ of $\approx 6^{\circ}$C every ~ millisecond (1/1200 s) at the time when a proton is released at the matrix side. Figure 4b illustrates the variations of $\Delta T$ in response to the three main "waves" or "blasts" of thermogeneration: an initial ~ $10^{-14}$–$10^{-13}$ s chemiosmotic (if uncoupled to ATP synthesis) and thermal bath wave; followed by a femtosecond protonation wave (vertical line, at the time resolution of the plot) resulting in $\Delta T \approx 4\ ^{\circ}$C (calculated from Eq. (6) by excluding the hydration term); and ending with a picosecond hydration step peaking at its completion before thermalization. We took 5 ps as a reasonable thermalization time since the beginning of these multistep heat bursts. The inset in Figure 4b is a more realistic singularity-free rendering of the variation in $\Delta T$ over a few picoseconds.

**Concluding remarks**

For the whole mitochondrion, proton leak governs heat generation since other heat generating mechanisms (e.g., protonation, solvation) cancel on average. Around ATP synthase, over time scales $\gtrapprox$ milliseconds, the sum of the $\Delta G$'s of protonation and solvation nearly cancel since the process is endergonic at the intermembrane gap side and exergonic at the matrix side.

"Gas-phase" protons are actively transported from the complexes of the ETC into the inter-membrane gap where they are protonated and hydrated (generating heat locally). By contrast, the uptake of $H^+$ in the ultimate reaction of the ETC – namely, $2H^+ + \frac{1}{2}O_2 + 2e^- \rightleftharpoons H_2O$ – requires the reverse sequence of reactions (i.e., dehydration and deprotonation) and hence will use energy locally. There are other processes that, as well, involve the near cancellation of the total free energy of the two opposing sequences of reactions.

The low thermal conductance of ATP synthase reflects the underlying *mechanism* of proton translocation. Proton translocation associated with Boyer's rotary mechanism, modeled as an axle-vane system, is limited by friction and torsion. In this manner, the $10^5$ gap is bridged, so to speak, and we now predict temperature differences of ≈ 6°C similar to that reported on the basis of MTY fluorescence measurements (≈ 10°C). This modeling of ATP synthase may have reconciled a millionfold disagreement between theory and experiment.

However, what do the experiments of Chrétien and colleagues measure (see also Outstanding questions)? We propose that they measure a time and ensemble statistical superposition of fluorescence spectra from MTY molecules located near the sites of protonation and hydration of the emerging protons at the matrix side. The continuous signal may be the results of a temporal and ensemble incoherent superposition of Avogadro's number of such spikes (number of protons/unit time × number of ATP synthases in the sample × number of nearby MTY molecules). Meanwhile, theoreticians are correct in that such a temperature gradient is impossible at a steady-state. *Thus, it appears that the Hot Mitochondrion Paradox is resolved once the pieces of the jigsaw – namely, the timescales, the mechanisms of heat generation and conductance, and the temporal and ensemble averaging – are assembled to complete the picture*. This picture is one of an inner mitochondrial membrane flanked by oscillating temperature differences both in the temporal domain and along its 2D surface.

A two-compartment thermodynamic system can perform work if the compartments exhibit a temperature difference. This concept (i.e., of the mitochondrion as a heat engine) has been proposed in a series of important papers by Muller [57–60]. In these papers, Muller elucidates the aspects of ATP synthase's operation as a molecular heat engine machine. Consistent with this line of work, it appears that this heat-engine-like behavior applies during ~ 1,000 ps bursts/s.

Voltage-gated ionic channels not infrequently operate *via* rotational mechanisms during ion translocation [61,62]. It is not unreasonable, perhaps, to conceive of these channels as ratchet engines as well. As such, these channels would exhibit an extremely low (non-canonical) thermal conductance. This is consistent with the proposition that a temperature difference of "tens of degrees" exist between the extremities of ion channels [63].

The idea of having temperature spikes is not in itself new. Rajagopal and colleagues demonstrate that inducible proton leak can lead to transient temperature gradients on much longer timescales (in seconds) [64]. Here, much shorter temperature spikes are shown to reconcile experiment and theory and hence may be a resolution of the hot mitochondrion paradox.

**Acknowledgements**

The authors thank Professor Thanh-Tung Nguyen-Dang (*Université Laval*) and Professor Lou Massa (*Hunter College, City University of New York*) for helpful discussions. The Natural Sciences and Engineering Research Council of Canada (NSERC), the Canada Foundation for Innovation (CFI), Laval University, and Mount Saint Vincent University are acknowledged for funding.

**Glossary**

**$10^5$ Gap:** a disagreement in the literature between experimentalists and theorists where the former report intracellular temperature differences in the order of one degree (or even ten degrees) while the latter places a theoretical limit at only $10^{-5}$ K over the distances typical of a living cell.

**ATP synthase:** a rotatory molecular machine enzyme that exist in mitochondria, in chloroplasts, and in bacteria and which is responsible for chemiosmotic coupling; that is, harnessing the electrochemical gradient into the formation of pyrophosphate bonds in ATP.

**Axle-vane engine:** a special case of the ratchet engine. It is a thermodynamic system comprising two separate temperature baths, at different temperatures, that are connected only through an axle terminated by vanes, one vane in each of the two systems. The engine transmits heat though the mechanical motion of the vanes as they are exposed to different thermal fluctuations at the two termini.

**Paradox:** an apparent inconsistency that results from a misunderstanding of existing facts or from facts that are omitted or missing.

**Ratchet engine:** a device whereby the motion is the result of the random Brownian thermal fluctuations after (mechanical, chemical, electrical, etc.) rectification. This engine was first imagined by Richard Feynman.

**Thermalization:** a reflection of the Second Law of thermodynamics whereby the average local differences in temperatures are equalized with time to maximize entropy.

**Box 1.**

**The Mitochondrion and ATP Synthase**

Mitochondria are cell organelles that are much smaller than the cell that contains them. The mitochondrion has a double-membrane structure: an outer smooth and highly permeable phospholipid bilayer surrounding a small intermembrane space, which encloses a highly invaginated and selective second phospholipid bilayer membrane. That latter (inner) membrane contains the mitochondrial matrix where the reactions of the tricarboxylic acid (TCA) (or Krebs) cycle occur.

The food we eat converges on the central "highway" of intermediary metabolism; that is, glycolysis. Glycolysis is very inefficient and is the main energy supplier for anaerobic metabolism, as it occurs, for instance, in highly dividing cells - such as cancer cells - that do not have time for the TCA cycle and the following ETC. Under normal aerobic conditions, one glucose ($C_6H_{12}O_6$) yields two equivalents of pyruvate (Pyr, $C_3H_3O_3^-$), which are transported into the mitochondrion where they are, each, "burned" into three $CO_2$ (per equivalent of Pyr) and, in the process, reduce the coenzymes $NAD^+$ and FAD. This happens in the TCA cycle in the matrix. The resulting reduced coenzymes, NADH and $FADH_2$, deliver their electrons and their accompanying protons, now each heading their separate ways, to Complexes I and II of the inner membrane (**Figure I**). The electrons travel through a cascade of coupled redox reactions of increasing reduction potential (all exergonic) until reaching the highest reducing agent in this series; namely hydrogen, which reacts with molecular (triplet) oxygen that comes from respiration. The reduction of respiratory $O_2$ leads to the formation of metabolic water (**Figure I**), but since this reaction is rarely accomplished stoichiometrically, it invariably leads to the formation of harmful peroxides and reactive oxygen species contributing to aging and degenerative diseases, mutations, lipid peroxidation, *etc*.

Meanwhile, the energy released from this cascade of exergonic reactions (ETC) is used to pump the protons ($H^+$) against the chemical and electric gradients maintained by metabolism across the inner membrane. This pumping maintains these gradients, which were first suggested to be the coupling mechanism for ATP synthesis by Peter Mitchell in the 1960s [65], work that was recognized by his 1978 Nobel Prize in Chemistry. The protons then find their way back, funneling down the gradients through ATP synthase, to re-enter the matrix releasing Gibbs energy.

This energy is mainly captured by the ATP synthase rotor system in the form of ATP. This opinion article argues, also, that this is the process responsible for the observed temperature gradient that is deemed impossible to understand under steady-state conditions (see text).

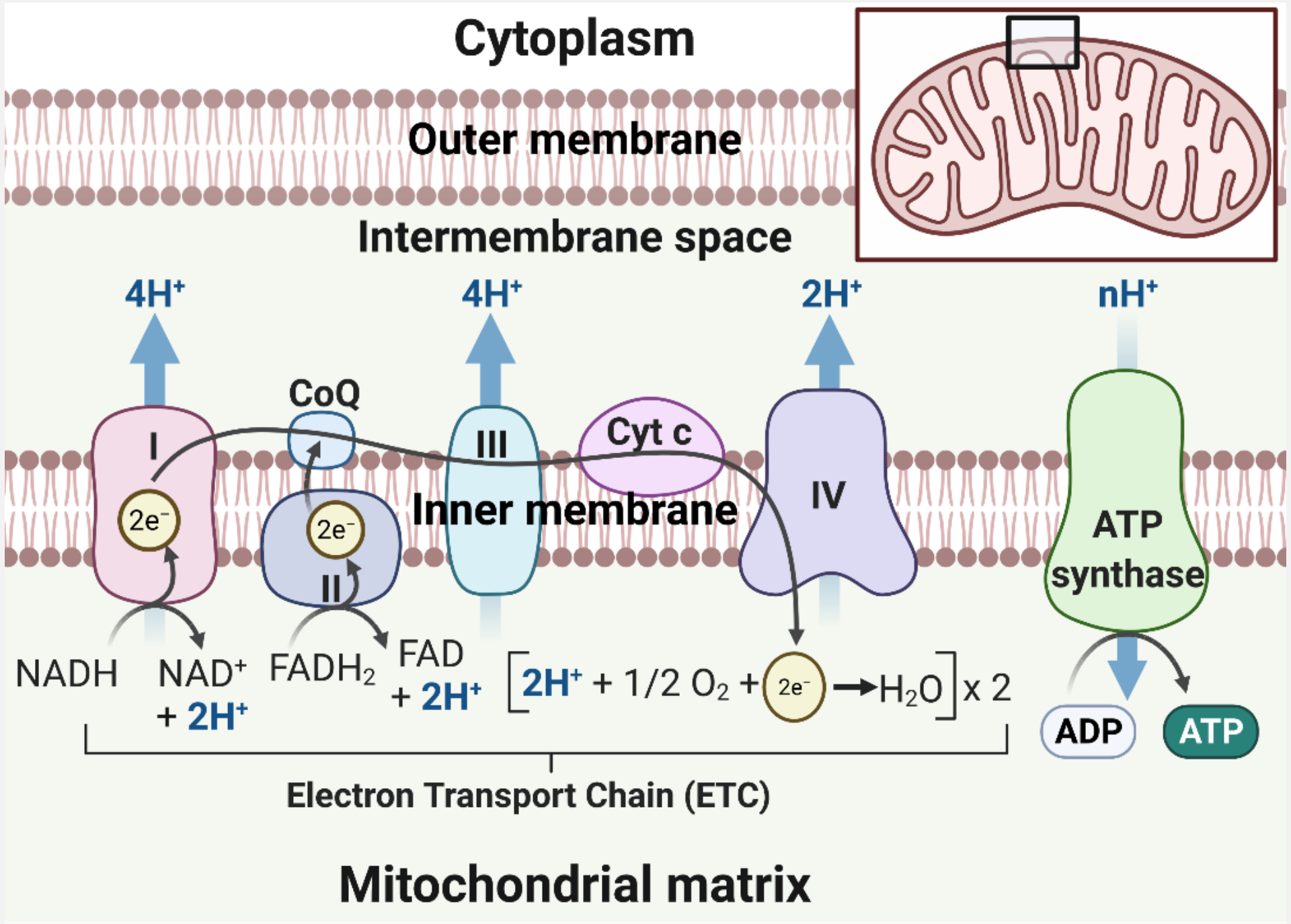

**Figure I:** A diagram showing the double-membrane structure of the mitochondrion with the principal enzymes and cofactors involved in the electron transport chain (ETC) and the oxidative phosphorylation of ADP into ATP. The ATP synthase enzyme captures the energy associated with the $H^+$ gradient released as each proton crosses the inner membrane to the matrix. This is also a mechanism for heat generation in the eukaryotic cell. Created with BioRender.com.

**Box 2.**

**Mitochondrial Molecular Thermometry using Florescent Dyes**

The fluorescence of rhodamine dyes (**Figure I**) has several desirable photophysical properties [66–69]. One of these is the pH insensitivity of the fluorescence intensity of this family of dyes [70]. A recent structure-activity-relationship study by Andresen and colleagues [71] confirmed the pH insensitivity of several rosamine scaffold-based fluorescent dyes as long as they lack the ionisable hydroxyl group attached to the phenyl ring [71]. Furthermore, rosamines, due to the lack of this carboxylic group, do not cyclize into lactones in solution as their rhodamine counterparts.

Ahn and colleagues [70] created a combinatorial library based on the rosamine scaffold in which the 2' carboxylic group in rhodamines that restricts the rotation of the 9-phenyl ring is removed. In this combinatorial approach, the two groups collected by the shaded squares in **Figure I** (*middle*) are varied independently [70]. Variations in *Group I* (the 3H-xanthene core) are affected by substitution(s) and by replacing the oxygen atom with other atoms or groups ("X"). *Group II* (the phenyl ring at position 9 of the xanthene nucleus) is varied by attaching different substituents to the ring. In this way, Ahn and colleagues created a library of 12 variations of *Group I* (which they label alphabetically from A to L) and 33 variations of *Group II*. The number of possible dies made of 12 variations of *Group I* and 33 of *Group II* is 12×33 = 396 dyes.

An important pH-insensitive rosamine dye is the combination I31 of xanthene scaffold "I" and phenyl substituent "31". Dye I31 is also called MTY [3] (**Figure I**) due to its affinity to mitochondria. This compound's pH insensitivity renders it particularly suited for probing the temperatures in the highly variable pH environment of the mitochondrion where pH gradients as high as 1 pH over a few nanometers can routinely be encountered.

Arai and co-workers [3] have shown that the fluorescence intensity of MTY correlates linearly with its ambient temperature *in vitro* and that the fluorescence, in addition to being pH insensitive, is also negligibly affected by ionic strength, external electric fields of moderate strengths, and the concentration of the dye itself [2].

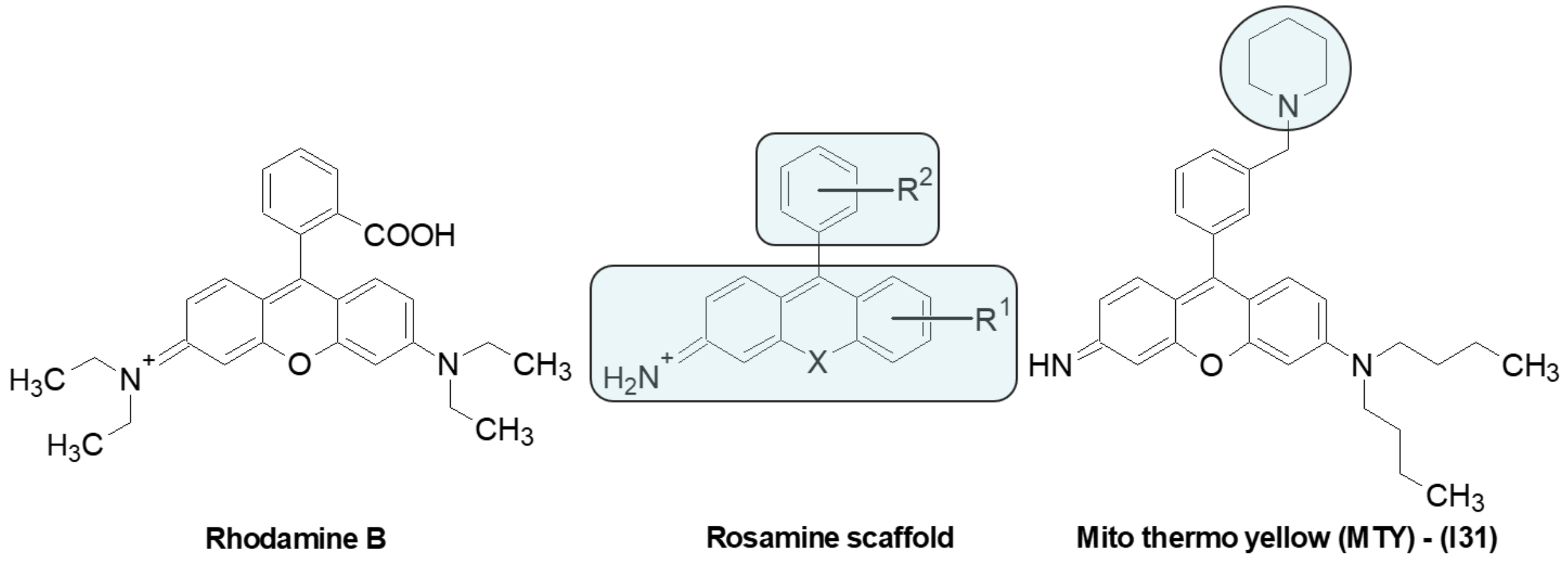

**Figure I:** Chemical structures of representatives of the rhodamine/rosamine class of molecular fluorescent thermometers. ***Left***: structure of a typical rhodamine dye. ***Middle***: a rosamine scaffold (decarboxylated rhodamine) with the two variable groups indicated by the shading. ***Right***: the structure of Mito thermo yellow (MTY) (I31) showing the chromophore in a circle.

By monitoring the fluorescence intensity using MTY as an *in situ* nano-thermometer, Chrétien and coworkers [2] suggested that mitochondrial temperatures, in human embryonic kidney (HEK) 293 cells and primary skin fibroblasts, are 10$^o$C higher than their ambient temperature maintained at 38$^o$C. This temperature difference disappeared on treatment with respiratory inhibitors, further lending credence to their proposal.

***A correction:***

The IUPAC nomenclature of MTY (dye I31) is N,N-dibutyl-3-imino-9-{3-[(piperidin-1-yl)methyl]phenyl}-3H-xanthen-6-amine (formula: $C_{33}H_{41}N_3O$, MW ≈ 495.7 Da). This correct structure of MTY (**Figure I**) is incorrectly displayed on two governmental websites where the piperidine ring and the phenyl ring are directly linked without the intervening methylene (formula: $C_{32}H_{39}N_3O$, MW ≈ 481.7 Da)[iii, iv].

**Box 3.**

**Fourier's Law of Heat Transfer**

According to the second law of thermodynamics, heat flows from hot regions to cold regions if there is a temperature difference within a system. Thermodynamics is concerned with the exchanged energy, equilibrium states, equilibrium temperatures, *etc.* irrespective of the underlying mechanisms of heat transfer.

Heat transfer can occur by conduction (our concern here), convection, or radiation. Heat transferred by conduction in solids occurs via the intermediacy of phonons; that is, collective lattice vibrational mode excitations. In liquids, conduction occurs primarily via collisions and the diffusion of molecules brought about by Brownian motion. The randomness of Brownian motion brings about thermalization; that is, the averaging out of the temperature over the system (with much larger dimensions than those of the composing molecules) in line with the second law. Heat is transmitted through a nonperiodic mesoscopic semirigid body such as a large molecular complex or a large protein through the excitation of its normal vibrational modes.

Fourier's law is an empirical expression relating the rate of heat transfer to the geometry and thermal conductivity of the material and the temperature gradient:

$$P = -\kappa A \nabla T, \tag{I}$$

in which $P$, $\kappa$, $A$, and $\nabla T$ are the power (heat transfer per unit time), the thermal conductivity coefficient, the area crossed perpendicularly by the heat flow vector, and the temperature gradient, respectively. The negative sign in this equation means that the direction of the power flux – that is, the integral of the power that crosses $A$ perpendicularly – points down the gradient as required by the second law. Eq. (I) strictly describes a steady-state situation, or one that can be approximated as a steady state, whereby the changes in temperature and rate of heat flow over time are ignored.

**Figure Ia** displays a thin spherical shell of thickness ($r_2$ - $r_1$) that could represent a mitochondrial membrane (inner, outer, or their conjunction with the gap). The Fourier law applied to this spherical system appears at the bottom of the left part of the figure. The right part **(b)** of this figure is the temperature profile for a homogenous sphere (with only radial dependence) obtained from the full-fledged heat equation for a steady state ($\frac{\partial T}{\partial t} = 0$):

$$\kappa \frac{1}{r^2} \frac{\partial}{\partial r} r^2 \frac{\partial T}{\partial r} + s = 0, \qquad \text{(II)}$$

where $s = 3P/(4\pi r_0^3)$ is the spherically averaged power density over the volume of this idealized (spherical) mitochondrion of radius $r_0$. The figure displays temperature as a function of the radial distance for a spherical mitochondrion when $P$ = 1 pW.

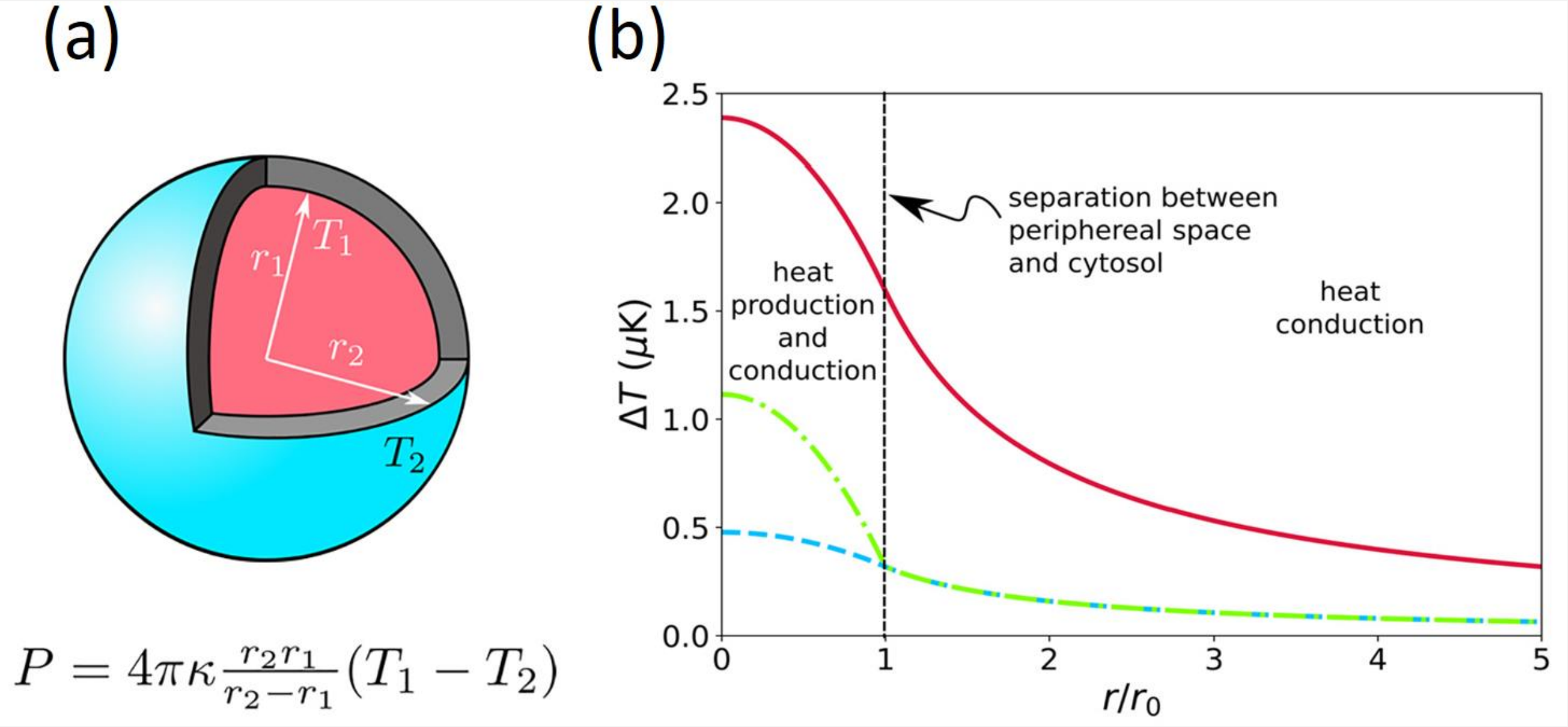

**Figure I:** Depictions of a spherical model of a mitochondrion and the distribution of temperature from its center to the outer cytoplasm, passing by its "effective" membrane structure. **(a)** An idealized representation of a spherical mitochondrion with a spherical-shell membrane of radius ($r_2$ – $r_1$). The expression of the Fourier law for this system is shown at the bottom. **(b)** Temperature difference profile obtained from the heat equation (II) for this spherical mitochondrion model assuming a spherically averaged power of 1 pW. The differently colored curves are calculated for different thermal conductivities; the reader is referred to the original paper by Macherel and colleagues [12] for details. Reproduced, with permission, from Figures 3 and 4 of [12].

**Box 4.**

**Ending the Confusion of Units/Dimensions involving Rotational Motion**

Units and dimensions are an integral part of the scientific discourse. The literature is full of confusion in rotational motion regarding the fate of angular units such as radians. Angles are comonly measured in degrees, but the SI unit is the radian. By definition, *the radian is the plane angle at the center of a circle subtended by two radii such that the length of the arc is equal to the radius*. A full circle has $2\pi$ radians. An angle is dimensionless since it is defined by reverse trigonometric functions as a ratio of two lengths; for example, $\theta = sin^{-1}\left(\frac{opposite}{hypotenuse}\right)$. There is no contradiction in being dimensionless and still being a unit of measurement. Often in physics and chemistry texts, radians disappear from equations by a sort of magic.

This endemic problem has been addressed by Edward S. Oberhofer of the *University of North Carolina at Charlotte* in an important paper [46] that appears to have been largely forgotten – a "hidden gem", so to speak. Oberhofer offers the following careful definitions:

(1) The commonly used (wrong) one, with SI units in brackets,

$$\theta(\text{rad}) = \frac{\text{arc of length } (S)\ (\text{m})}{\text{radius of length } (R)\ (\text{m})}, \qquad \text{(I)}$$

where the unit "rad" appears to have popped out of nowhere; and

(2) the less frequently used (correct) definition,

$$\theta(\text{rad}) = \frac{\text{arc of length } (S)\ (\text{m})}{\text{arc length per radian of angle subtended } (R)\ (\text{m/rad})}, \qquad \text{(II)}$$

where the denominator is *not* simply the radius, but "*R* is rather the *arc length per radian* of angle subtended, that arc length being equal to the radius" [46]. This is the only way the units balance. Once this subtle but *crucial* distinction is observed, everything else falls into place and all units balance in equations [46].

As an example, suppose we have a cylinder that rotates at an angular speed of 4 rad/s and, say, its radius is 2 m, then the linear speed of its circumference is calculated usually as $v = \omega R$ = (4 rad/s)×(2 m) = 8 rad m/s, but then simply quoted as m/s (as it should). Notice that, to get the correct answer, *R* must be set to 2 m/rad [46]. Oberhofer offers a few illustrative examples (see

Table 1 in [46]), from which we quote the example relevent to this opinion article; that is, the moment of inertia.

In rotational motion, the moment of inertia plays the role of mass in linear motion. Meanwhile, angular velocity substitutes linear velocity and torque replaces froce. The moment of inertia $I$ is defined in many textbooks as being (*mass of the rotating point)* × (*length of the radius or rotation*)$^2$ which has the SI units kg.m$^2$. While this will not represent any calculational problems, it does make the units on both sides of some equations inconsistent. In the present paper, this subtly has been taken into consideration and, instead of the unit of $\mathrm{kg.m^2}$, as in most physics textbooks, here we adopt the unit $\frac{\mathrm{kg.m^2}}{\mathrm{rad^2}}$ necessary to balance the units properly as described by Oberhofer [46].

**Resources**

i https://www.uniprot.org/uniprot/A0A287A9I8

ii https://www.uniprot.org/uniprot/Q9CR84

iii https://pubchem.ncbi.nlm.nih.gov/compound/Mito-thermo-yellow

iv https://jglobal.jst.go.jp/en/detail?JGLOBAL_ID=201507001277137164&rel=1#%7B%22category%22%3A%220%22%2C%22keyword%22%3A%22mito%20thermo%20yellow%22%7D

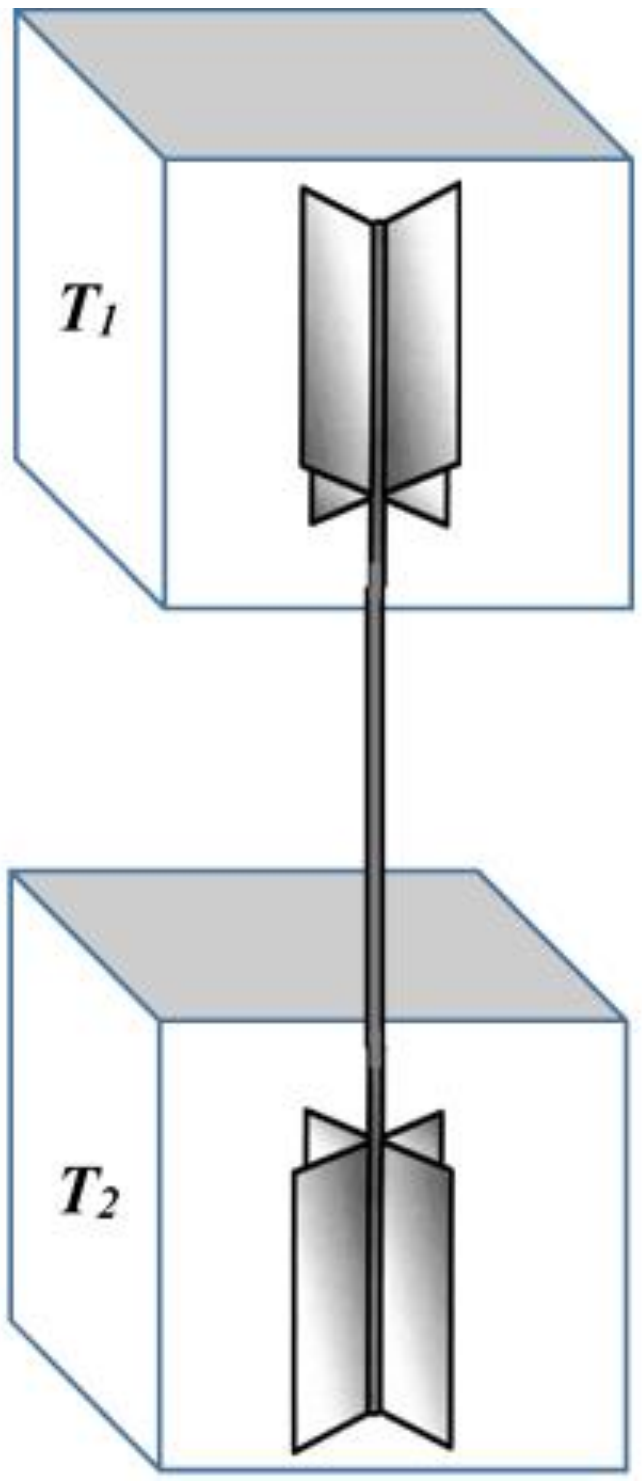

**Figure 1** Parrondo and Español's "axle-vane" modification of Feynman's ratchet engine.

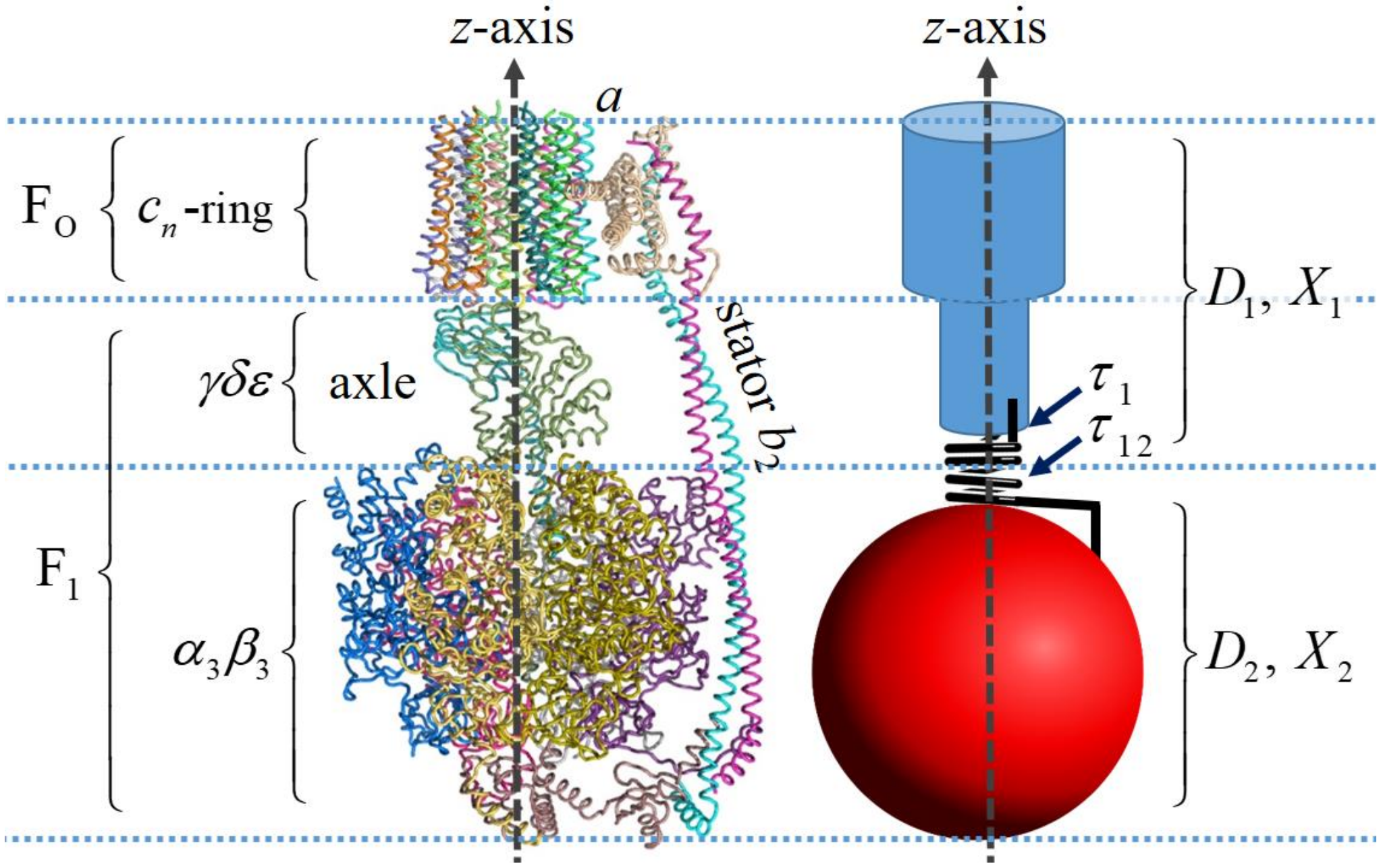

**Figure 2** Idealization of the rotary structure of ATP synthase as a tortionally controlled two-part system. ***Left*:** a model of ATP synthase showing the two principal regions ($F_O$ and $F_1$). ***Right*:** an idealization showing the spring-like connection of the $\gamma$-subunit to the $\alpha_3\beta_3$-subunit, with two different spring constants $\tau_1$ and $\tau_{12}$. This is a slightly modified version of that of Okazaki and Hummer [45]. While the axle is labelled $\gamma\delta\varepsilon$ in the diagram of ATP synthase (since it comprises three subunits), this is not the "axle" referred to in the "axle-vane engine" since the model to the right considers only the torsion of the connector between the $\gamma$-subunit and the $\alpha_3\beta_3$-subunit.

$$\left[\begin{array}{ll} H_3O^+(aq) \xrightleftharpoons{\text{de-solvation}} H_3O^+(g) + (H_2O_{solv})_n, & \Delta G^0 = +6.9\times10^{-19}\ \text{J/proton},\quad \tau_s = 10^{-12}\text{s},\ \Delta\tau_s \approx 8\times10^{-17}\text{s} \\ H_3O^+(g) \xrightleftharpoons{\text{de-protonation}} H^+(g) + H_2O(g), & \Delta G^0 = +1.3\times10^{-18}\ \text{J/proton},\quad \tau_s = 10^{-15}\text{s},\ \Delta\tau_s \approx 4\times10^{-17}\text{s} \\ \hline H_3O^+(aq) \xrightleftharpoons{\text{overall}} H^+(g) + H_2O(g) + (H_2O_{solv})_n, & \Delta G^0 = 2.0\times10^{-18}\ \text{J/proton} \end{array}\right]_{\text{Inter-membrane space}}$$

$$\left[ H^+_{\substack{\text{Inter-membrane}\\ \text{space}}}(g) \xrightleftharpoons{\text{proton translocation}} H^+_{\substack{\text{Matrix}\\ \text{just released}}}(g) \qquad \Delta G^0_{CO} = -3.2\times10^{-20}\ \text{J/proton},\ \Delta G^0_{TB} = -4.1\times10^{-21}\ \text{J/proton} \right]_{\text{Translocation}}$$

$$\left[\begin{array}{ll} H^+_{\substack{\text{Matrix}\\ \text{just released}}}(g) \xrightleftharpoons{\substack{\text{proton in transit}\\ \text{(coupled to ATP synthesis)}}} H^+_{\substack{\text{colliding}\\ \text{with } H_2O}}(g) & \Delta G^0_{TB} = -4.1\times10^{-21}\ \text{J/proton},\ \tau_s = 1.4\times10^{-13}\text{s},\ \Delta\tau_s \approx 1\times10^{-14}\text{s} \\ H^+_{\substack{\text{Matrix}\\ \text{just released}}}(g) \xrightleftharpoons{\substack{\text{proton in transit}\\ \text{(leak, uncoupled to ATP}\\ \text{synthesis)}}} H^+_{\substack{\text{colliding}\\ \text{with } H_2O}}(g) & \Delta G^0_{CO+TB} = -3.6\times10^{-20}\ \text{J/proton},\ \tau_s = 4.5\times10^{-14}\text{s},\ \Delta\tau_s \approx 1\times10^{-15}\text{s} \\ H^+(g) + H_2O(g) \xrightleftharpoons{\text{protonation}} H_3O^+(g), & \Delta G^0 = -1.3\times10^{-18}\ \text{J/proton},\quad \tau_s = 10^{-15}\text{s},\ \Delta\tau_s \approx 4\times10^{-17}\text{s} \\ H_3O^+(g) + (H_2O_{solv})_n \xrightleftharpoons{\text{solvation}} H_3O^+(aq) & \Delta G^0 = -6.9\times10^{-19}\ \text{J/proton},\quad \tau_s = 10^{-12}\text{s},\ \Delta\tau_s \approx 8\times10^{-17}\text{s} \\ \hline H^+(g) + H_2O(g) + (H_2O_{solv})_n \xrightleftharpoons{\substack{\text{overall}\\ \text{(coupled to ATP synthesis)}}} H_3O^+(aq), & \Delta G^0 = -2.0\times10^{-18}\ \text{J/proton} \\ H^+(g) + H_2O(g) + (H_2O_{solv})_n \xrightleftharpoons{\substack{\text{overall}\\ \text{(leak, uncoupled to ATP}\\ \text{synthesis)}}} H_3O^+(aq), & \Delta G^0 = -2.0\times10^{-18}\ \text{J/proton} \end{array}\right]_{\text{Matrix}}$$

**Figure 3** Endergonic processes (intermembrane gap side) and exergonic processes (matrix side) with/without coupling to ATP synthesis. The subscripts "CO" emphasizes "chemiosmotic" and "TB" thermal bath, while "CO+TB" means that both contribute to a given process. The top, middle, and bottom brackets collect, respectively, the reactions that occur in the intermembrane gap, the process of proton translocation through ATP synthase's $F_O$ subunit, and the reactions occuring in matrix on the proton's release. If the translocation step is coupled with ATP synthesis, only the TB energy is available for dissipation. (All values are rounded to, at most, two significant figures, yet it is probably safer to consider all values as indicative of orders of magnitude.)

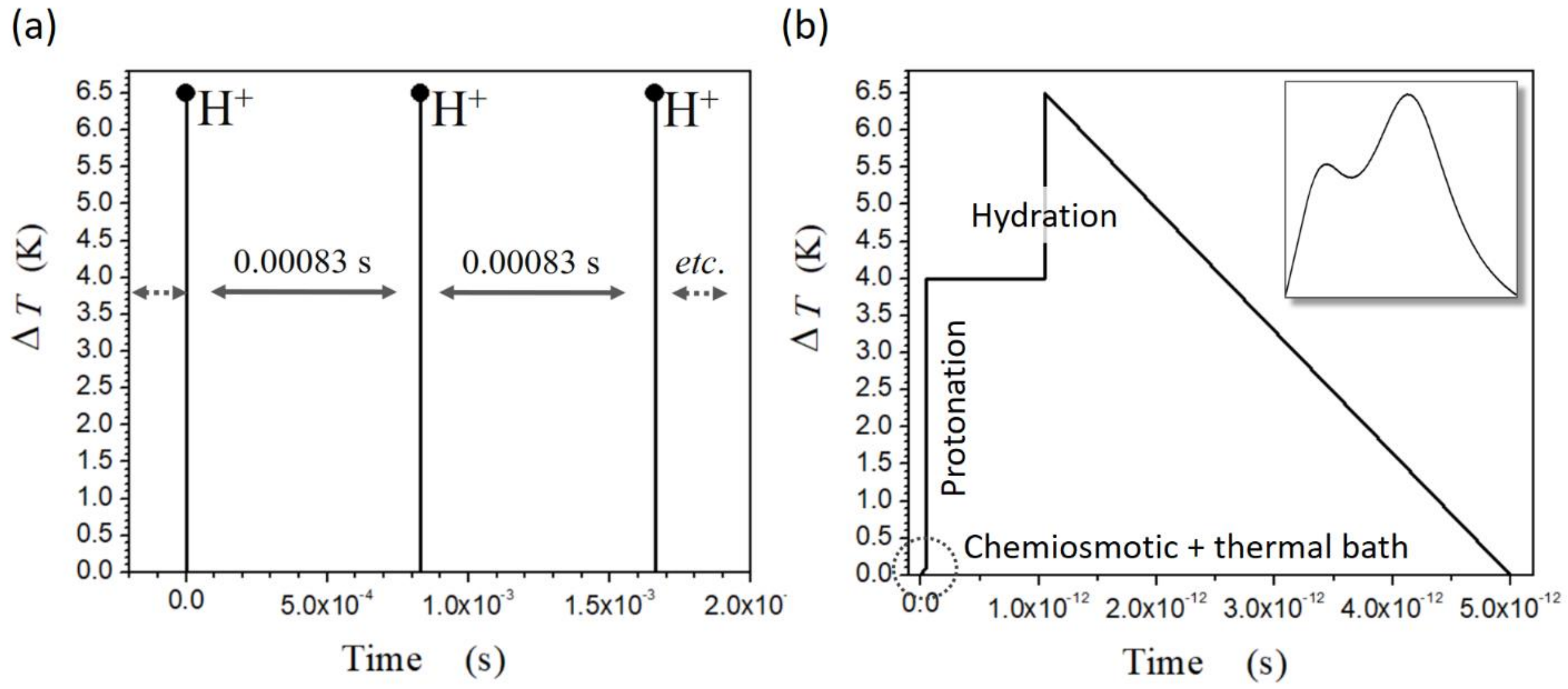

**Figure 4** Temporal dependence of the temperature difference ($\Delta T$) across the $F_O$ unit of ATP synthase at two timescales. **(a)** Profile of $\Delta T$ at millisecond tempral resolution. **(b)** An idealized plot of the response of $\Delta T$ to the three principal "waves" of heat production; namely, the (chemiosmotic) and thermal bath wave (at ~ $10^{-14}$ - $10^{-13}$ s timescale) followed by the steep protonation step (at ~ $10^{-15}$ s timescale – too narrow to have any but a vertical-linear representation), which overlaps with the slower hydration (~ $10^{-12}$ s) that peaks near its completion before the decay due to the dissipation of heat. Full thermalization is assumed to occur within 5 ps. The inset in **(b)** is a more realistic smoothing of the plot over ~ 5 ps.